\documentclass[aps,prl,numerical,superscriptaddress,showpacs,floatfix,reprint]{revtex4-1}
\usepackage{amsmath,amssymb,mathrsfs,bm} 
\usepackage{url}
\usepackage{hyperref}
\hypersetup{
colorlinks=true,
linkcolor=blue,
anchorcolor =red,
citecolor=blue,
filecolor = red,
urlcolor=blue,
pdfauthor=author}
\usepackage{txfonts}
\usepackage[utf8]{inputenc}
\usepackage{amsmath}
\usepackage{graphicx}
\def\av<#1>{\left\langle\,#1\,\right\rangle}
\def\ev<#1>{\left\langle\,#1\,\right\rangle_{\rm{ev}}}

\begin{document}


\title{Azimuthal Anisotropy Scaling for Identified Mesons and Baryons: Insights into \\Medium Transport Properties, Equation of State and Hadronic Re-scattering}

\author{ Roy~A.~Lacey}
\email[E-mail: ]{Roy.Lacey@Stonybrook.edu}
\affiliation{Department of Chemistry, 
Stony Brook University, \\
Stony Brook, NY, 11794-3400, USA}
%
%

\date{\today}
\begin{abstract}
Scaling functions for the centrality and transverse momentum dependence of \(v_2(p_T,\text{cent})\) and \(v_3(p_T,\text{cent})\) are constructed for identified mesons and baryons in Pb+Pb (\(\sqrt{s_{NN}}=2.76,\ 5.02\)~TeV), Xe+Xe (5.44~TeV), and Au+Au (0.2~TeV) collisions. These species-resolved functions capture the interplay of initial geometry, viscous attenuation, radial flow, partonic energy loss, and hadronic re-scattering across both flow- and quenching-dominated regimes. The systematic growth of the radial-flow parameter \(\zeta_{\rm rf}\) with multiplicity and beam energy provides direct empirical constraints on the equation of state (EOS) of hot QCD matter. The extracted parameters also yield differential constraints on the specific shear viscosity \(\eta/s\), the jet-quenching parameter \(\hat{q}\), and late-stage hadronic dynamics. LHC systems exhibit low \(\eta/s\), strong radial flow, and negligible re-scattering at high energy density, whereas Au+Au at RHIC energy shows even lower \(\eta/s\), weaker radial flow, finite re-scattering, and reduced energy density. The coexistence of strong radial flow at low \(p_T\) and significant jet suppression at high \(p_T\) emerges as a defining hallmark of QGP formation, establishing the framework as a quantitative probe of QGP transport properties and EOS stiffness.
\end{abstract}

%

\pacs{25.75.-q, 25.75.Dw, 25.75.Ld} 
\maketitle

%
High-energy heavy-ion collisions at the Large Hadron Collider (LHC) and Relativistic Heavy Ion Collider (RHIC) create the quark-gluon plasma (QGP), a state of matter that replicates conditions microseconds after the Big Bang~\cite{Shuryak:1983ni,Bass:1998vz,Satz:2000bn}. Determining the QGP’s transport coefficients and equation of state (EOS) is essential for understanding its evolution and thermodynamics.

A key probe of these properties is azimuthal anisotropy, the medium’s response to initial spatial asymmetries. At low transverse momentum ($p_T$) it arises from collective flow, while at higher $p_T$ it encodes path-length–dependent partonic energy loss (jet quenching). Anisotropy is quantified by Fourier coefficients $V_n$~\cite{Bilandzic:2010jr,Luzum:2011mm,Teaney:2012ke}:
\begin{equation}
V_n \equiv v_n e^{in\Psi_n} = \langle e^{in\phi} \rangle,
\label{Vndef}
\end{equation}
where $v_n$ is the anisotropy magnitude, $\Psi_n$ the symmetry-plane angle, and $\langle \cdot \rangle$ denotes an average over the single-particle azimuthal distribution. Experimentally, $v_n$ is often extracted with the Scalar Product (SP) method~\cite{Adler:2002pu,Voloshin:2008dg}, which correlates the unit flow vector of a particle of interest, $k$, with an event-wide Q-vector from reference particles:
\[
\textbf{u}_{n,k} = \exp(in\varphi_k), \qquad
\textbf{Q}_n = \sum w_i e^{in \varphi_i},
\]
where $\varphi_k$ and $\varphi_i$ are azimuthal angles and $w_i$ the particle weight. A pseudorapidity gap $(\Delta\eta)$ between the particle of interest and reference sub-events suppresses non-flow correlations. The $v_n$ coefficients are then given by
\begin{equation}
v_n \{\mathrm{SP}\} = \frac{\left\langle \langle \textbf{u}_{n,k}{\textbf{Q}^*_n} \rangle \right\rangle}{\sqrt{\frac{\langle \textbf{Q}_n \textbf{Q}^{A*}_n \rangle \langle \textbf{Q}_n \textbf{Q}^{B*}_n \rangle}{\langle \textbf{Q}^A_n \textbf{Q}^{B*}_n \rangle}}},
\end{equation}
with ${}^*$ denoting complex conjugation. Single brackets $\langle \cdot \rangle$ indicate event averages, and double brackets $\langle \langle \cdot \rangle \rangle$ averages over particles within and across events. The denominator corrects for finite event-plane resolution via sub-event correlations.

Extensive studies~\cite{Song:2010mg,Alver:2008zza,Alver:2010rt,Ollitrault:2009ie,Dusling:2009df,Lacey:2010fe,Shen:2011eg,Niemi:2012aj,Fu:2015wba,Andres:2015ara,STAR:2022gki,STAR:2018fpo,ALICE:2016kpq,Qiu:2011iv,Adare:2011tg,Magdy:2018itt,Adamczyk:2016gfs,STAR:2015rxv,Adamczyk:2015obl,Adamczyk:2016exq,Adam:2019woz,Gardim:2014tya,Holopainen:2010gz,Qin:2010pf,Qiu:2011iv,Gale:2012rq,JET:2013cls,Liu:2018hjh,PHENIX:2001hpc,STAR:2002ggv,Zhang:2008fh,ALICE:2013dpt,Mehtar-Tani:2013pia,Qin:2015srf} show that the anisotropy coefficients \(v_n(p_T,\text{cent})\) reflect the combined influence of eccentricities \((\varepsilon_n)\) and their fluctuations, viscous attenuation, radial flow, partonic energy loss, and hadronic re-scattering. For \(p_T \lesssim 4{-}5\)~GeV, \(v_n\) is dominated by collective flow, while at higher \(p_T\) jet quenching \emph{is the primary driver}~\cite{Lacey:2024fpb,Liu:2018hjh,Gardim:2014tya}.

Recent work~\cite{Lacey:2024fpb} introduced anisotropy \emph{scaling functions} for charged particles, providing unified descriptions of \(v_{2,3}(p_T,\text{cent})\) across broad \(p_T\) and centrality ranges~\cite{ALICE:2010suc,ATLAS:2011ah,ATLAS:2012at,CMS:2012xss,CMS:2012zex,ALICE:2011ab,ATLAS:2013xzf,CMS:2013wjq}. These functions constrained the specific shear viscosity \(\eta/s\), the jet-quenching parameter \(\hat{q}\), and the spectrum and fluctuations of \(\varepsilon_n\). However, averaging over species masks characteristic variations driven by mass, baryon number, and hadronic cross section. Such variations carry essential information on radial flow and re-scattering, and provide additional constraints on transport coefficients and the EOS.

Species-resolved measurements recover this missing information. Since \(\varepsilon_n\) and the transport coefficients characterize the common medium, species-dependent differences in \(v_n\) mainly reflect particle mass, baryon number, and hadronic cross section. At low to intermediate \(p_T\), radial flow boosts heavier hadrons—especially baryons—to higher transverse momenta, producing a characteristic blue shift of their spectra. This underlies the well-known mass and baryon-number ordering of \(v_n(p_T)\), whose magnitude grows for a stiffer EOS with larger pressure-to-energy-density ratio. Thus, variations masked in averages directly reveal how radial flow and the EOS shape the collective response.  

Additional discrimination comes from comparing hadrons with small hadronic cross sections (e.g., \(K^+\), \(\phi\), \(\Xi^{\pm}\), \(\Omega^{\pm}\)) to those with large ones (e.g., \(\pi^\pm\), \(p\), \(\bar{p}\)). The former largely preserve partonic flow and the radial-flow–driven blue shift, while the latter are more strongly modified by late-stage re-scattering. Such species-resolved contrasts sharpen sensitivity to transport coefficients and the EOS, motivating the extension to species-resolved scaling functions.

Building on this basis, the present study extends the anisotropy scaling methodology~\cite{Lacey:2024fpb} to identified mesons (\(\pi^{\pm}, K^{\pm}, K^0_S, \phi\)) and baryons (\(p, d, {^3\text{He}}, \Lambda^0, \Xi^-, \Omega^-\)). The goal is to establish species-resolved scaling functions for \(v_2(p_T,\text{cent})\) and \(v_3(p_T,\text{cent})\) that capture the distinct roles of radial flow, viscous attenuation, and hadronic re-scattering across system size, centrality, and beam energy. The framework incorporates key elements: eccentricities \(\varepsilon_n\) and their fluctuations, the dimensionless system size \((\mathbb{R} \propto RT)\), and a viscous correction \((\delta_f)\) to the thermal distribution. While \(\eta/s\), \(\hat{q}\), and the initial geometry are common to all species in a given event class, their manifestations in \(v_n(p_T)\) are modulated by species-dependent mass and hadronic cross section. This enables differential constraints on the initial conditions, radial flow strength, and the interplay of viscous damping and hadronic re-scattering.

Event-by-event correlations between \(\langle p_T \rangle\) and \(v_n\) fluctuations have been proposed as indirect probes of radial flow~\cite{Parida:2024ckk,Schenke:2020uqq}, but the measured fluctuations arise from multiple sources, including initial-geometry fluctuations, viscous attenuation, hadronization, hadronic re-scattering, and partonic energy loss. Their combined influence can obscure radial-flow signatures and complicate efforts to isolate its role in QGP evolution. In contrast, the species-resolved scaling functions developed here provide a more differential approach, disentangling the mechanisms that shape anisotropic flow without conflating them with fluctuation-driven correlations. By systematically incorporating eccentricity, transport coefficients, jet quenching, and late-stage interactions, the framework yields direct constraints on the medium’s collective response.

For species-resolved scaling, the kaon scaling function in ultra-central (uc) 5.02~TeV Pb+Pb serves as the universal baseline: kaons have intermediate mass, small hadronic cross section, and high-precision $v_{2,3}(p_T)$; uc geometry further limits baseline bias. Unlike charged-hadron scaling—which averages over species and conflates baryon number, mass, and cross section—the kaon anchor preserves species information and exposes physics-driven offsets. After applying $\varepsilon_n$, the size scale $\mathbb{R}$, and the viscous correction $\delta f$, \textit{interspecies} differences shrink but persist. The \emph{species-resolved} scaling functions therefore \emph{retain} the charged-hadron attenuation parameter $\beta=k_\beta\beta_0$ and \emph{add} late-stage variables that isolate re-scattering ($\zeta_{\rm hs}$, entering mesons via $\zeta_m=1-\zeta_{\rm hs}$) and radial flow ($\zeta_{\rm rf}$, entering baryons via $\zeta_b=(1-\zeta_{\rm rf})^{|n_B|}$). Normalization exponents ensure cross-system and inter-harmonic consistency.

Building on the preceding formulation, the meson $v_2$ scaling relation is
\begin{multline}
\frac{v_2(p_T,\mathrm{uc})}{\varepsilon_2(\mathrm{uc})}
e^{\tfrac{2\beta_0}{\mathbb{R}_{\rm uc}}(2+\kappa p_T^2)}
=
e^{\alpha\,\tfrac{2\beta_0}{\mathbb{R}_{\rm uc}}(2+\kappa p_T^2)\,\zeta_M^{(2)}}
\\[-2pt]
\times
\left(\frac{v_2'(p_T)}{\varepsilon_2'}\right)^{\zeta_m}
e^{\tfrac{2\zeta_m\beta}{\mathbb{R}_{\rm uc}}
\left(\tfrac{\mathbb{R}_{\rm uc}}{\mathbb{R}'}-1\right)(2+\kappa p_T^2)} ,
\label{eq:v2_scaling_mesons}
\end{multline}
where primes denote the \emph{comparison} system or centrality and $\mathbb{R}'$ its size factor.
Here, $\beta=k_\beta\beta_0$, $\delta f=\kappa p_T^2$, and $\mathbb{R}$ decreases with centrality.
The normalization exponent is $\zeta_M^{(2)}=[\,\zeta_m/k_\beta+\gamma_{32}\,]$, where $\gamma_{32}$ is a geometry-only normalization tied to inter-system differences in $(\varepsilon_3/\varepsilon_2)$ and estimated from those ratios (discussed below).
The construction aims to achieve a cross-system collapse with a shared normalization; $\gamma_{32}$ absorbs these $(\varepsilon_3/\varepsilon_2)$ differences while leaving the attenuation/size factors ($\propto\,2\beta_0/\mathbb{R}$ and $\mathbb{R}_{\rm uc}/\mathbb{R}'$) and the $p_T$ slope set by $k_\beta$ unchanged. For uc, $\alpha=1$; for near-uc (0–5\%), $\alpha\approx 0.5$ (bin-averaged); for non-uc, $\alpha$ is absorbed into the normalization and $(2{+}\kappa p_T^2)\!\to\!1$.

The viscous correction \(\delta_f = \kappa p_T^2\) (quadratic ansatz) governs attenuation across the full \(p_T\) range: capturing viscous effects at low and intermediate \(p_T\) and regulating scaling in the jet-quenching regime at high \(p_T\). To maintain continuity, \(\delta_f\) is fixed above a threshold \(p_T^{\rm thresh} \sim 4.5\) GeV/\(c\), where partonic energy loss dominates \cite{Lacey:2024fpb}. This treatment provides unified constraints on both \(\eta/s\) and \(\hat{q}\).

An independent constraint is provided by the mapping from \(v_3\) to \(v_2\), expressed as
\begin{multline}
\frac{v_2(p_T,\text{uc})}{\varepsilon_2(\text{uc})}\cdot
e^{\tfrac{2\beta_0}{\mathbb{R}_{\rm uc}}(2+\kappa p_T^2)} =
e^{\alpha\cdot \tfrac{2\beta_0}{\mathbb{R}_{\rm uc}}(2+\kappa p_T^2)\cdot \zeta_M^{(2\leftarrow3)}} \\
\times
\left(\frac{v_3'(p_T)}{\varepsilon_3'}\right)^{\tfrac{2}{3}\zeta_m}
e^{\tfrac{2\zeta_m\beta}{\mathbb{R}_{\rm uc}}
\left(\frac{\mathbb{R}_{\rm uc}}{\mathbb{R}'}-1\right)(3+\kappa p_T^2)},
\label{eq:v2_from_v3_mesons}
\end{multline}
where \(\zeta_M^{(2 \leftarrow 3)} = \tfrac{2}{3}\left[1- (\zeta_m/k_{\beta} +\gamma_{32}) \right]\).

Together, Eqs.~\ref{eq:v2_scaling_mesons} and \ref{eq:v2_from_v3_mesons} impose complementary intra- and inter-harmonic constraints, providing a stringent, predictive test of meson scaling across beam energies, system sizes, and centralities.

Analogously for baryons, the $v_2$ scaling relation is
\begin{multline}
\frac{v_2(p_T, \text{uc})}{\varepsilon_2(\text{uc})} \cdot 
e^{ \tfrac{2 \beta_0}{\mathbb{R}_{\rm uc}} (2 + \kappa p_T^2) } = 
e^{ \alpha \tfrac{2 \beta_0}{\mathbb{R}_{\rm uc}} 
(2 + \kappa p_T^2) \zeta_B^{(2)} } \\
\times \left( \frac{v_2'(p_T)}{\varepsilon_2'} \right)^{\zeta_b} \cdot 
e^{ \tfrac{2 \zeta_b \beta}{\mathbb{R}_{\rm uc}} 
\left( \tfrac{\mathbb{R}_{\rm uc}}{\mathbb{R}'} - 1 \right)(2 + \kappa p_T^2) }.
\label{eq:v2_scaling_baryons}
\end{multline}
Here $\zeta_b = (1-\zeta_{\rm rf})^{|n_B|}$ encodes an \emph{empirical} baryon-number dependence of the radial-flow normalization, with $|n_B|$ ensuring equal treatment of baryons and antibaryons. The normalization exponent is defined as $\zeta_B^{(2)} = \left[\zeta_b - \big({1}/{k_{\beta}} - \gamma_{32} \big)\right]$, where $\gamma_{32}$ is the same parameter used for mesons.

An independent constraint comes from the inter-harmonic mapping $v_3 \!\to\! v_2$, with the scaling relation
\begin{multline}
\frac{v_2(p_T, \text{uc})}{\varepsilon_2(\text{uc})} \cdot 
e^{ \tfrac{2\beta_0}{\mathbb{R}_{\rm uc}} (2 + \kappa p_T^2) } = 
e^{ \alpha \tfrac{2\beta_0}{\mathbb{R}_{\rm uc}} (2 + \kappa p_T^2) \zeta_B^{(2 \leftarrow 3)} } \\
\times \left( \frac{v_3'(p_T)}{\varepsilon_3'} \right)^{\tfrac{2}{3}\zeta_b} \cdot 
e^{ \tfrac{2 \zeta_b \beta}{\mathbb{R}_{\rm uc}} 
\left( \tfrac{\mathbb{R}_{\rm uc}}{\mathbb{R}'} - 1 \right)(3 + \kappa p_T^2) }.
\label{eq:v2_from_v3_baryons}
\end{multline}

\begin{figure*}[tbh]
    \centering
    \includegraphics[clip,width=0.75\linewidth]{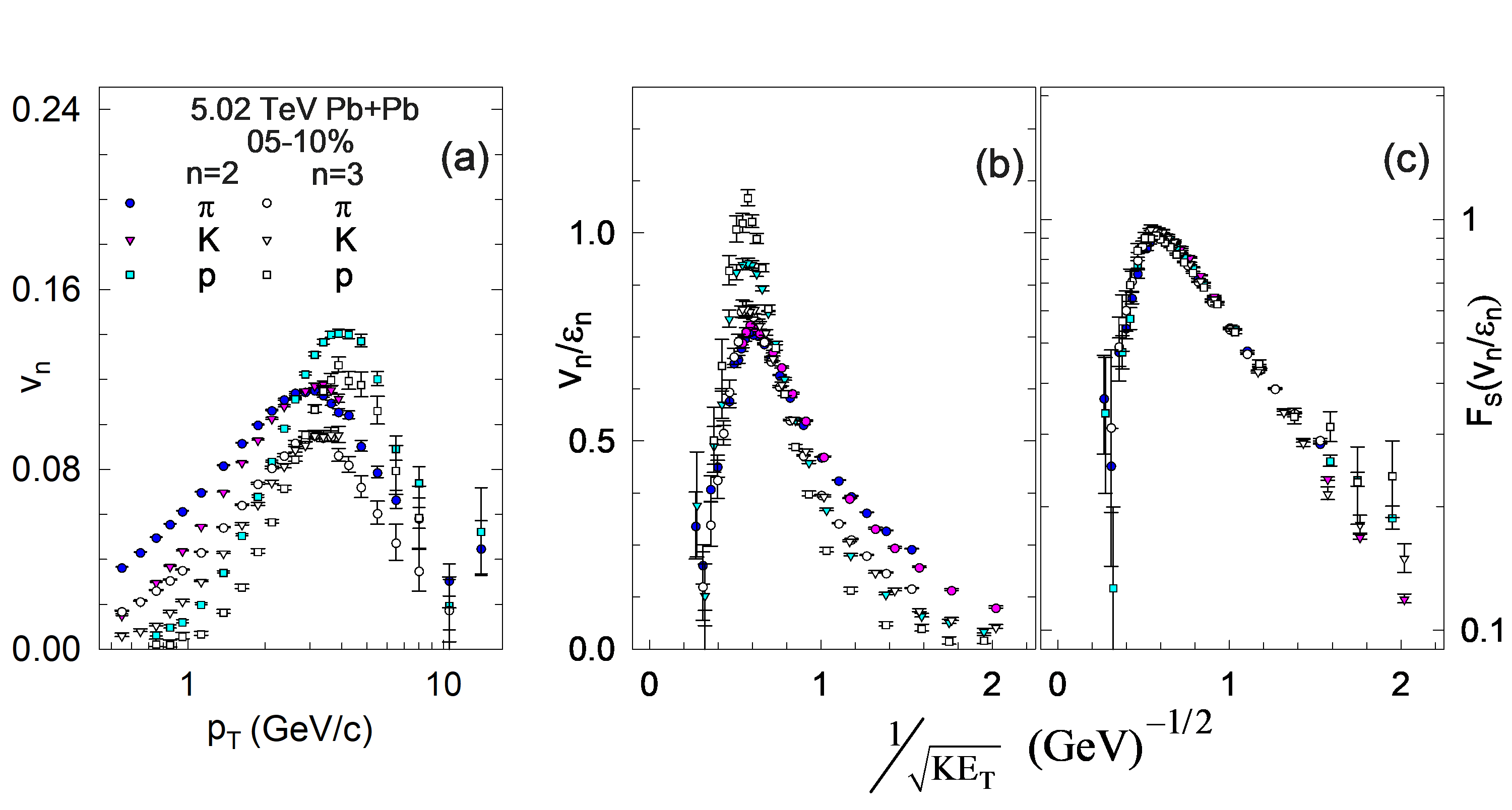}
    \vskip -0.3 cm
    \caption{(Color online)
Scaling of $v_2$ and $v_3$ in 5--10\% central Pb+Pb collisions at $\sqrt{s_{NN}} = 5.02$ TeV. 
Panel (a): unscaled $v_2(p_T)$ and $v_3(p_T)$ for $\pi$, $K$, and $p$, showing mass and baryon-number ordering. 
Panels (b,c): eccentricity-scaled and fully scaled results vs.\ $1/\sqrt{KE_T}$. 
Eccentricity scaling reduces but does not remove species separation (b), while full scaling yields near-universal behavior across flow- (large $1/\sqrt{KE_T}$, low $p_T$) and quenching-dominated (small $1/\sqrt{KE_T}$, high $p_T$) regions. 
Data: ALICE~\cite{Zhu:2019twz,ALICE:2022zks}.
}
    \label{fig1}
\end{figure*}
Here $\zeta_B^{(2 \leftarrow 3)} = \left[\zeta_b - \big( \tfrac{5}{3k_\beta} - \gamma_{32} \big)\right]$, 
providing the normalization for the mapping.  

Together, Eqs.~\ref{eq:v2_scaling_baryons} and \ref{eq:v2_from_v3_baryons} supply complementary intra- and inter-harmonic constraints for baryons. Combined with the meson results, they form a unified, species-resolved framework disentangling initial geometry, viscous damping, radial flow, partonic energy loss, and hadronic re-scattering. Simultaneous agreement across systems, centralities, and $p_T$ then tests the scaling formalism and constrains $\eta/s$, $\hat{q}$, and EOS-sensitive observables in nuclear collisions.

The framework also encodes sensitivity to initial-state geometry—particularly nuclear deformation—through its dependence on $\varepsilon_n$. Eccentricity-scaling provides a natural handle on geometry-driven effects, enabling both intra-harmonic ($v_n/\varepsilon_n$) and inter-harmonic ($v_3/\varepsilon_3 \!\to\! v_2/\varepsilon_2$) scaling across centralities, systems, and energies. The inter-harmonic mapping is especially geometry-sensitive: $v_3$ is largely fluctuation-driven, whereas $v_2$ has strong geometric contributions. Thus, the centrality dependence of $v_3/\varepsilon_3 \!\to\! v_2/\varepsilon_2$ scaling provides a stringent constraint on deformation. In $^{129}$Xe, quadrupole deformation enhances $\varepsilon_2$ in uc and near-uc collisions, modifying the geometric component of $v_2$. The observed scaling collapse in Xe+Xe supports modeling $^{129}$Xe with $\beta_2 \approx 0.18\pm 0.01$~\cite{Lacey:2024fpb}. For nuclei with octupole deformation, where $\varepsilon_3$ acquires a stronger geometric contribution, the centrality dependence of $v_3/\varepsilon_3 \!\to\! v_2/\varepsilon_2$ provides a complementary probe of higher-order shape. Because transport, radial flow, and hadronic re-scattering are simultaneously constrained, the extraction of nuclear shape parameters is robust and avoids the assumptions that often complicate $v_n$ analyses. These results highlight the broader utility of the scaling framework as a diagnostic of both QGP transport and nuclear structure.

\begin{figure*}[tbh]
    \centering
    \includegraphics[clip,width=0.75\linewidth]{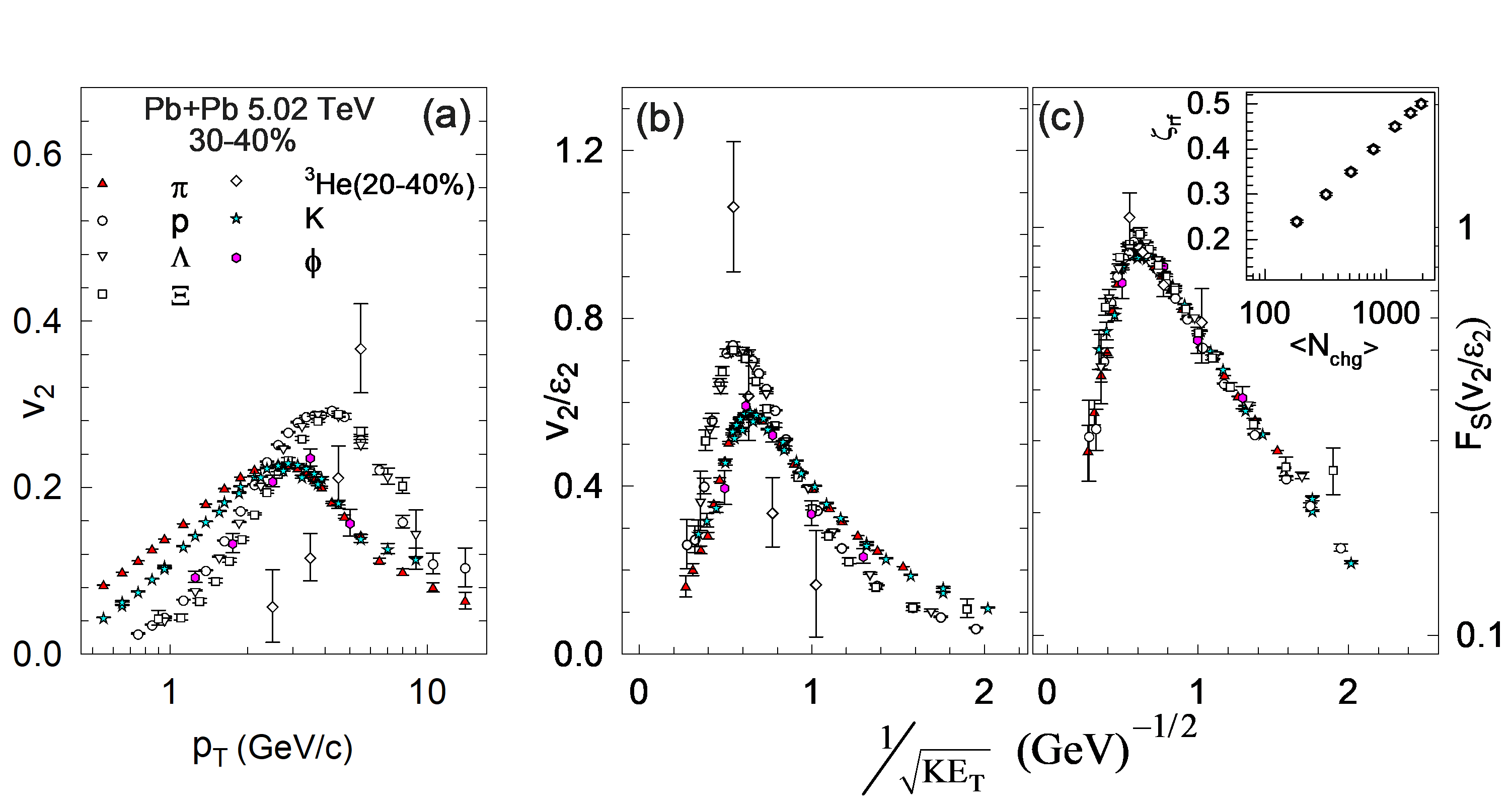}
    \vskip -0.3 cm
    \caption{(Color online)
Scaling of $v_2$ in 30--40\% central Pb+Pb collisions at $\sqrt{s_{NN}} = 5.02$ TeV. 
Panel (a): measured $v_2(p_T)$ for mesons ($\pi$, $K$, $K^0_S$, $\phi$) and baryons ($p$, $\Lambda^0$, $\Xi$, $^{3}$He). 
Panels (b,c): eccentricity-scaled and fully scaled results vs.\ $1/\sqrt{KE_T}$. 
Inset (c): multiplicity dependence of $\zeta_{\rm rf}$. 
Data: ALICE~\cite{Zhu:2019twz,ALICE:2020chv,ALICE:2022zks}.
}
    \label{fig2}
\end{figure*}

\begin{figure*}[th]
    \centering
    \includegraphics[clip,width=0.75\linewidth]{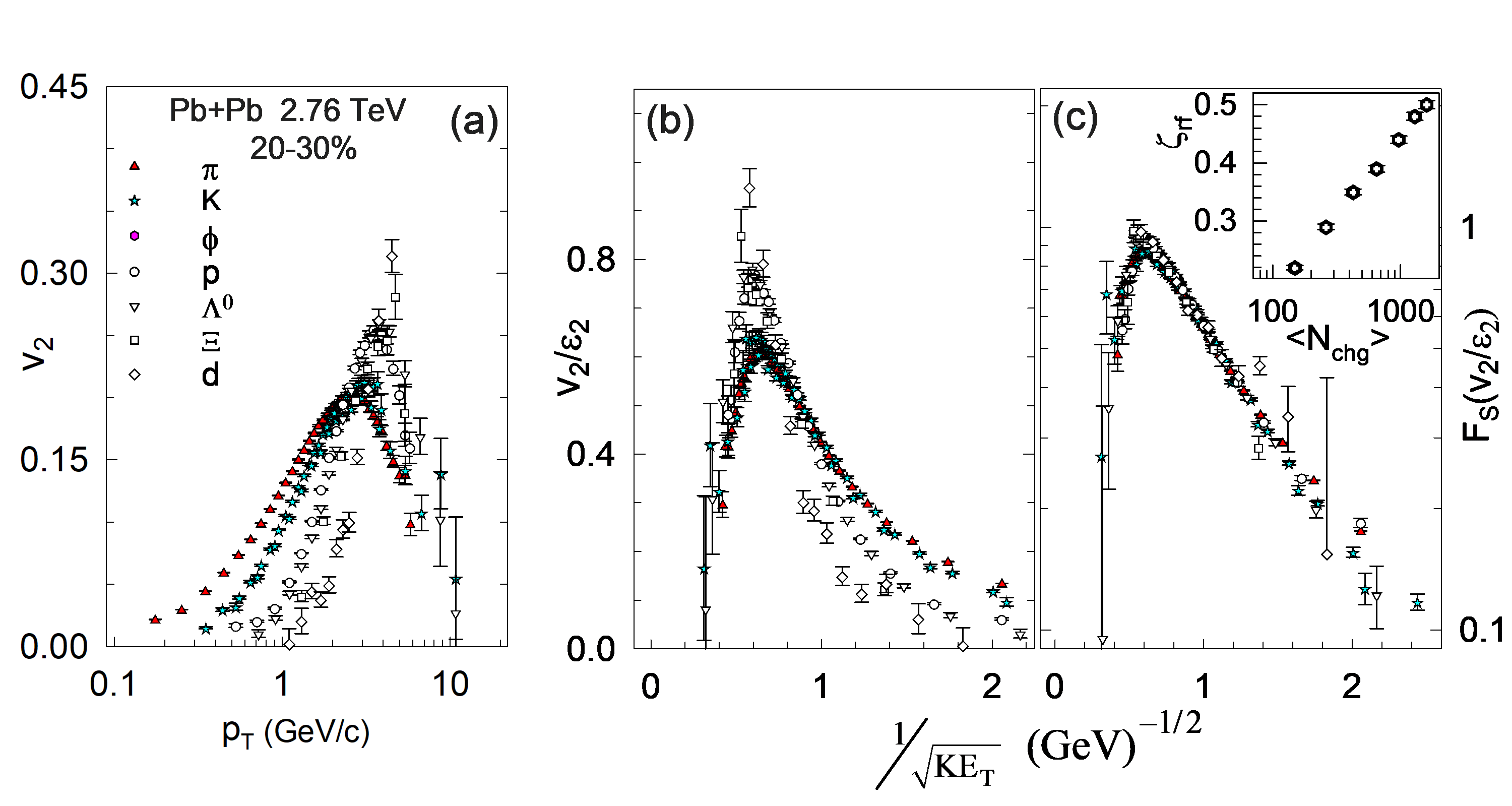}
    \vskip -0.3 cm
   \caption{(Color online)
Scaling of $v_2$ in 20--30\% central Pb+Pb collisions at $\sqrt{s_{NN}} = 2.76$ TeV. 
Panel (a): measured $v_2(p_T)$. 
Panels (b,c): eccentricity-scaled and fully scaled results vs.\ $1/\sqrt{KE_T}$, following Fig.~\ref{fig2}. 
Inset (c): multiplicity dependence of $\zeta_{\rm rf}$. 
Data: ALICE~\cite{ALICE:2014wao,ALICE:2016cti,ALICE:2017nuf,ALICE:2018lao}.
}
    \label{fig3}
\end{figure*}

\begin{figure*}[th]
    \centering
    \includegraphics[clip,width=0.75\linewidth]{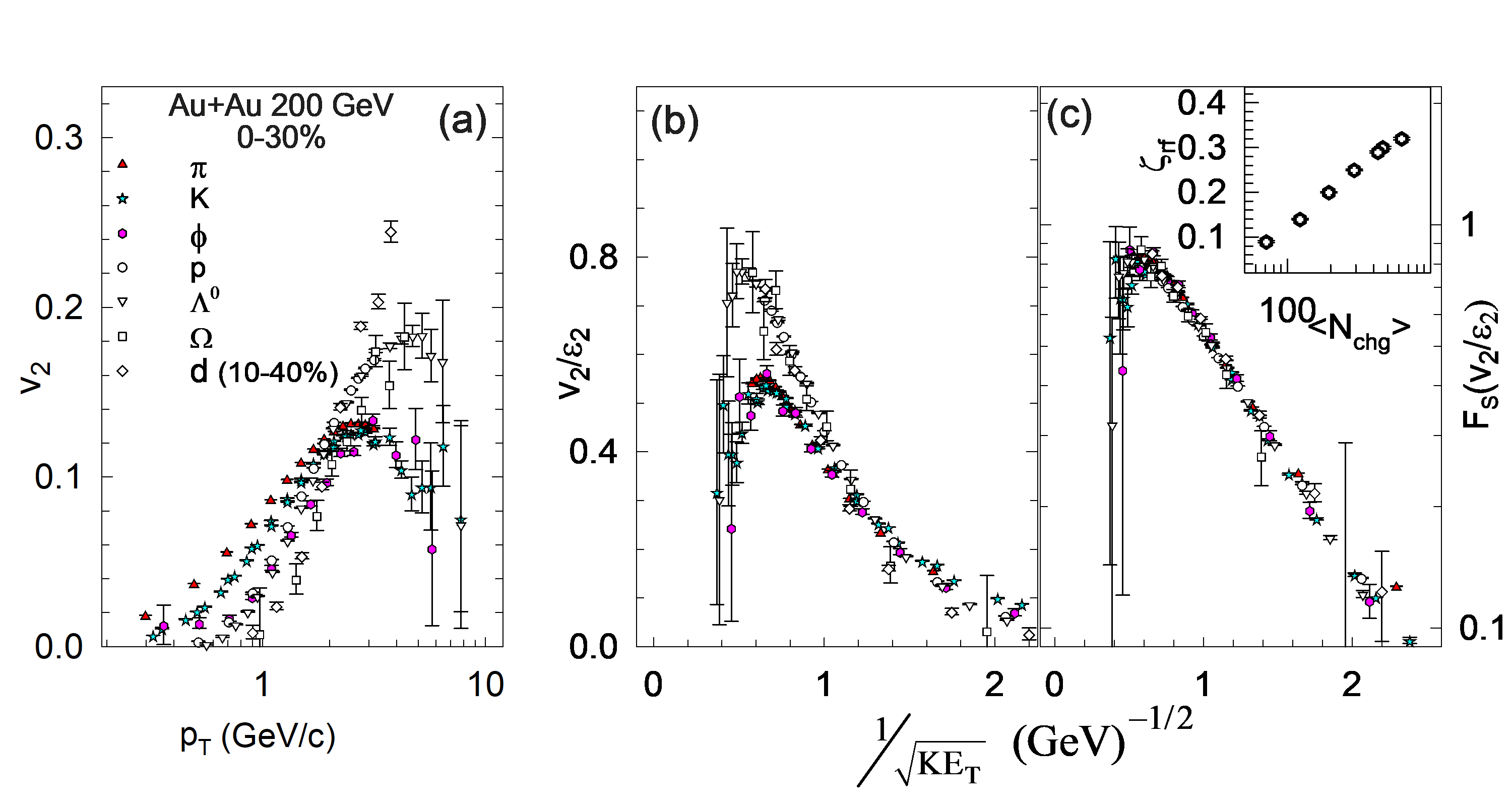}
    \vskip -0.3 cm
 \caption{(Color online)
Scaling of $v_2$ in 0--30\% central Au+Au collisions at $\sqrt{s_{NN}} = 0.20$ TeV. 
Panel (a): measured $v_2(p_T)$. 
Panels (b,c): eccentricity-scaled and fully scaled results vs.\ $1/\sqrt{KE_T}$. 
Full scaling (c) achieves near-universal behavior across mesons ($\pi$, $K$, $K^0_S$, $\phi$) and baryons ($p$, $\Lambda^0$, $\Omega$, $d$), spanning collective to quenching-dominated regimes. 
Inset (c): multiplicity dependence of $\zeta_{\rm rf}$. 
Data: PHENIX and STAR~\cite{PHENIX:2014uik,STAR:2015gge,STAR:2022ncy}.
}
    \label{fig4}
\end{figure*}

The characteristic transverse scale $\mathbb{R} \propto \langle N_{\text{chg}} \rangle^{1/3}$, used to normalize viscous attenuation and system-size–dependent prefactors in Eqs.~\ref{eq:v2_scaling_mesons}--\ref{eq:v2_from_v3_baryons}, was determined from centrality-dependent charged-particle multiplicities $\langle N_{\text{chg}} \rangle_{|\eta|\le 0.5}$ obtained from pseudorapidity densities~\cite{ALICE:2010mlf,ALICE:2015juo,ALICE:2018cpu,CMS:2019gzk,Lacey:2016hqy}. Here $\mathbb{R}$ acts as a proxy for the system’s spatial extent and initial energy density. The viscous correction was taken as $\delta f = \kappa p_T^2$, with $\kappa = 0.17~(\mathrm{GeV}/c)^{-2}$ from Ref.~\cite{Liu:2018hjh}.  

The data set comprises $v_2(p_T, \text{cent})$ and $v_3(p_T, \text{cent})$ for identified species in Pb+Pb at $\sqrt{s_{NN}} = 2.76$~TeV~\cite{ALICE:2014wao,ALICE:2016cti,ALICE:2017nuf,ALICE:2018lao} and 5.02~TeV~\cite{Zhu:2019twz,ALICE:2020chv,ALICE:2022zks}, Xe+Xe at 5.44~TeV~\cite{ALICE:2021ibz}, and Au+Au at 0.2~TeV~\cite{PHENIX:2014uik,STAR:2015gge,STAR:2022ncy}, reported by ALICE, PHENIX, and STAR. Values are averaged over particles and antiparticles, as differences lie within uncertainties.  

Initial-state eccentricities $\varepsilon_{2,3}(\text{cent})$ were obtained with the Monte Carlo quark–Glauber (MC-qGlauber) model~\cite{Liu:2018hjh}, which extends the MC-Glauber framework~\cite{Miller:2007ri,PHOBOS:2006dbo} by including quark substructure, realistic nucleon spatial distributions, finite nucleon size, and beam-energy–dependent inelastic cross sections via a smoothed wounding profile. Simulations were performed for Pb+Pb (5.02, 2.76~TeV), Xe+Xe (5.44~TeV), and Au+Au (0.20~TeV). Nuclear shapes were modeled with deformed Woods–Saxon densities; for $^{129}$Xe a quadrupole deformation $\beta_2=0.18 \pm 0.01$ was used, consistent with the Xe+Xe scaling analysis~\cite{Lacey:2024fpb}, while Au and Pb were taken as spherical. Parameter variations indicate $\mathcal{O}(2\text{–}3)\%$ uncertainties on $\varepsilon_n$. The resulting $\varepsilon_{2,3}(\text{cent})$ enter the species-resolved scaling relations in Eqs.~\ref{eq:v2_scaling_mesons}--\ref{eq:v2_from_v3_baryons}.

In near-spherical A{+}A, $(\varepsilon_3/\varepsilon_2)$ is largest in uc events (showing near system-independence) and \emph{decreases} toward peripheral, approaching a broad mid–peripheral plateau~\cite{Lacey:2024fpb}; smaller systems (e.g., Xe{+}Xe) yield systematically larger $(\varepsilon_3/\varepsilon_2)$, especially in the plateau region. The geometry-only normalization used in the scaling is defined as $\gamma_{32}\equiv\ln[(\varepsilon_3/\varepsilon_2)_{\rm ref}/(\varepsilon_3/\varepsilon_2)_{\rm sys}]$, evaluated in the same regime; it tracks these inter-system differences and sets the normalization offset.

Species-resolved scaling functions were constructed from differential $v_2(p_T, \text{cent})$ and $v_3(p_T, \text{cent})$ using the meson and baryon relations in Eqs.~\ref{eq:v2_scaling_mesons}--\ref{eq:v2_from_v3_baryons}. Meson coefficients followed Eqs.~\ref{eq:v2_scaling_mesons}--\ref{eq:v2_from_v3_mesons}, baryon coefficients Eqs.~\ref{eq:v2_scaling_baryons}--\ref{eq:v2_from_v3_baryons}, and all species were mapped to a common reference curve $\mathrm{F_S}(v_n/\varepsilon_n)$. This preserves the dual-constraint structure—enforcing intra-harmonic ($v_2$) and inter-harmonic ($v_3 \!\to\! v_2$) consistency—and yields a species-agnostic constraint on viscous transport, radial flow, partonic energy loss, and hadronic re-scattering.  

For cross-species comparison, the transverse kinetic energy ${\rm KE}_T = m_T - m_0$ was used as the scaling variable, with $m_T = \sqrt{p_T^2 + m_0^2}$ the transverse mass and $m_0$ the rest mass. This choice reduces trivial kinematic mass effects, emphasizes collective flow, and supports consistent scaling between mesons and baryons.

Figure~\ref{fig1} illustrates the scaling procedure for 5--10\% central Pb+Pb collisions at $\sqrt{s_{NN}} = 5.02$~TeV, with $1/\sqrt{{\rm KE}_T}$ on the abscissa in panels (b) and (c) to separate flow- and quenching-dominated regions~\cite{Dokshitzer:2001zm,Lacey:2010fe}. Panel (a) shows the raw $v_2(p_T)$ and $v_3(p_T)$, including mass ordering and the sizable $v_2$--$v_3$ separation for each species.  

Panel (b) demonstrates that applying eccentricity- and ${\rm KE}_T$-scaling, with $\varepsilon_n$ as the geometric seeds of anisotropy, improves agreement across species—most clearly between pions and kaons. Residual blue shifts between protons and mesons, and the systematic $v_2$--$v_3$ offset, reflect differences in hadronic cross sections, species-dependent radial flow, and the differing viscous and quenching sensitivities of the two harmonics.  

Panel (c) presents the final scaling function, where all species collapse onto a common curve spanning both domains. The extracted parameters, $\beta_0 = 0.88$, $k_{\beta} = 1.0$ (so $\beta = \beta_0$), $\zeta_{\rm rf} = 0.48$, $\zeta_{\rm hs} = 0.00$, and $\gamma_{32} = 0$, indicate anisotropies dominated by QGP physics: viscous transport ($\eta/s$), partonic energy loss ($\hat{q}$), and strong radial flow. The latter links directly to high opacity, since both radial flow and jet quenching arise from dense QGP conditions. The vanishing hadronic term confirms negligible late-stage interactions.  

The collapse across mesons and baryons of different mass, quark content, and cross section underscores the versatility of the framework. The scaling function $\mathrm{F_S}(v_n/\varepsilon_n)$ cleanly disentangles geometric seeds ($\varepsilon_n$) from the QGP response, yielding stringent, species-agnostic constraints on transport coefficients and collective expansion.

Figure~\ref{fig2} shows a similar analysis for $v_2(p_T)$ in 30--40\% central Pb+Pb collisions at $\sqrt{s_{NN}} = 5.02$~TeV, including mesons ($\pi, K, K^0_S, \phi$) and baryons ($p, \Lambda^0, \Xi^-, {^3\text{He}}$). Panel (a) illustrates the species-dependent $v_2(p_T)$ trends, analogous to Fig.~\ref{fig1}(a). Panel (b) shows partial convergence after eccentricity- and ${\rm KE}_T$-scaling: mesons collapse onto a common meson curve, while $n_B=1$ baryons align on a separate baryon curve, blue shifted relative to mesons. ${^3\text{He}}$ exhibits a larger residual blue shift than $n_B=1$ baryons, signaling enhanced radial flow tied to its higher baryon number. The remaining splittings reflect the combined effects of radial flow and viscous attenuation, which depend on particle type and centrality.  

Panel (c) presents the final scaling function $\mathrm{F_S}(v_n/\varepsilon_n)$, with data collapsing onto a single curve for parameters $k_{\beta} = 1.0$, $\zeta_{\rm hs} = 0.00$, $\zeta_{\rm rf} = 0.35$, and $\gamma_{32} = 0$. These values confirm negligible hadronic re-scattering and show that anisotropies are governed by initial eccentricities, viscous transport ($\eta/s$), partonic energy loss ($\hat{q}$), and species-dependent radial flow. The coexistence of strong radial flow at low $p_T$ and significant jet quenching at high $p_T$ reflects their shared origin in the medium’s high initial energy density, consistent with a strongly coupled QGP.  

The inclusion of ${^3\text{He}}$ ($|n_B| = 3$) provides a stringent test: its alignment with the reference scaling confirms the baryon-number dependence $\zeta_b = (1 - \zeta_{\rm rf})^{|n_B|}$. The smaller blue shift in mid-central collisions (30--40\%) reflects a reduction in radial flow relative to 5--10\% (cf. Fig.~\ref{fig1}), consistent with the decrease in $\zeta_{\rm rf}$ from $0.45$ to $0.35$. This reduction matches the lower multiplicity and energy density characteristic of mid-central collisions.

Validation of the scaling function was performed for 5.02~TeV Pb+Pb collisions across 0--1\% to 50--60\% centrality. Fixed parameters $\beta_0 = 0.88$, $k_{\beta}=1.0$ (so $\beta=\beta_0$), and $\zeta_{\rm hs} = 0.00$ were used, while $\zeta_{\rm rf}$—quantifying the species-specific radial-flow response—was varied with centrality. The extracted $\zeta_{\rm rf}$ values, shown in the Fig.~\ref{fig2} inset versus average multiplicity $\langle N_{\text{chg}} \rangle_{|\eta| \le 0.5}$, follow a logarithmic rise with $\langle N_{\text{chg}} \rangle$, reflecting the growth of radial flow with entropy, and hence energy, density.  

The stability of $k_{\beta}$, and $\zeta_{\rm hs}$ across centralities underscores the robustness of the framework and the reliability of the eccentricity spectrum. In contrast, the steady increase of $\zeta_{\rm rf}$ isolates the EOS-driven growth of radial flow with energy density.

State-of-the-art hydrodynamic models, such as IP-Glasma+MUSIC~\cite{Schenke:2010rr,Gale:2012rq}, successfully describe anisotropic flow—especially $v_2$ and $v_3$—at low transverse momentum ($p_T < 2.5$~GeV/$c$), where collective dynamics dominate. These models have constrained key QGP properties, including $\eta/s$ and the initial-state eccentricities, but their predictive power diminishes at higher $p_T$, where partonic energy loss and jet quenching prevail.  

The success of hydrodynamics at low $p_T$ is consistent with the observed scaling in this regime. Agreement with the species-resolved scaling function $\mathrm{F_S}(v_n/\varepsilon_n)$ provides a critical cross-check and reinforces confidence in the model assumptions. Importantly, the scaling functions span the full $p_T$ range—from collective to quenching-dominated—yielding a unified framework that disentangles and constrains $\eta/s$, $\hat{q}$, radial flow, and hadronic re-scattering. This broad reach enables extension into the intermediate- and high-$p_T$ regions, where energy-loss attenuation dominates and sensitivity to the temperature dependence of QGP transport coefficients is enhanced.

Robust species-resolved scaling functions—closely matching those for 5.02~TeV Pb+Pb—were also obtained for $v_2(p_T,\text{cent})$ and $v_3(p_T,\text{cent})$ in Xe+Xe at 5.44~TeV, Pb+Pb at 2.76~TeV, and Au+Au at 0.2~TeV. Figures~\ref{fig3} and \ref{fig4} show representative results for 20--30\% central Pb+Pb and 0--30\% central Au+Au, respectively. In each case, scaling functions were built with the system-dependent $k_{\beta}$, $\zeta_{\rm hs}$, and $\gamma_{32}$, while $\zeta_{\rm rf}$ varied with centrality. Relative to the 5.02~TeV Pb+Pb baseline ($k_{\beta}=1.0$), $k_{\beta}$ decreases by $\sim 5\%$ in 2.76~TeV Pb+Pb, increases by $\sim 2\%$ in 5.44~TeV Xe+Xe, and drops in 0.2~TeV Au+Au ($k_{\beta}=0.63$). Hadronic re-scattering is negligible in Pb+Pb and Xe+Xe ($\zeta_{\rm hs}=0$) but finite in Au+Au ($\zeta_{\rm hs}=0.08$). Likewise, $\gamma_{32}$ vanishes in Pb+Pb, is modestly negative in Xe+Xe ($-0.31$), and slightly positive in Au+Au ($0.05$).

The close agreement between scaling functions in Xe+Xe and Pb+Pb confirms that, at high energies, final-state anisotropies are dominated by QGP viscosity and radial flow. In both systems, $\zeta_{\rm rf}$ rises logarithmically with multiplicity, with comparable magnitudes at matched $\langle N_{\text{chg}}\rangle$. The vanishing $\zeta_{\rm hs}$ signals negligible hadronic re-scattering, while modest $k_\beta$ variations—a $\sim$5\% decrease at 2.76~TeV and $\sim$2\% increase at 5.44~TeV relative to the 5.02~TeV baseline ($k_\beta=1.0$)—indicate only minor changes in $\eta/s$ and $\hat{q}$ across the LHC energy range.  

In contrast, the lower $\beta$ and finite $\zeta_{\rm hs}$ in 0.2~TeV Au+Au reflect reduced partonic energy loss, diminished opacity, and non-negligible hadronic re-scattering—conditions consistent with a cooler medium and shorter QGP lifetime. The drop in $k_\beta$ relative to LHC values points to a reduced $\eta/s$, linked to the temperature dependence of shear viscosity near $T_c$. Meanwhile, weaker jet quenching and radial flow arise from the lower energy density, limiting both opacity and expansion. As shown in the Fig.~\ref{fig4} inset, $\zeta_{\rm rf}$ again grows logarithmically with multiplicity but remains systematically below Pb+Pb values at the same $\langle N_{\text{chg}} \rangle$, underscoring reduced collectivity at lower energies and the EOS sensitivity of the radial-flow response.

The continuity of species-resolved scaling functions from low-$p_T$ collectivity to high-$p_T$ energy loss provides quantitative constraints on both $\eta/s$ and $\hat{q}$ and supports the relation $\eta/s \propto T^3/\hat{q}$~\cite{Majumder:2007zh,JETSCAPE:2020mzn}, consistent with the inequality $\eta/s > T^3/\hat{q}$ implied by $\delta f$ viscous-correction formalisms~\cite{Dusling:2009df}. The coexistence of strong radial flow at low $p_T$ and jet suppression at high $p_T$—together with species-specific corrections from $\zeta_{\rm rf}$ and $\zeta_{\rm hs}$—is a hallmark of QGP formation and underscores the temperature sensitivity of its transport coefficients. The observed system-size scaling further highlights the role of $\mathbb{R} \propto \langle N_{\rm chg} \rangle^{1/3}$ and initial eccentricities $(\varepsilon_n)$ in shaping anisotropies, establishing a unified framework for constraining QGP transport, the equation of state, and dynamical evolution.

This study presents a species-resolved scaling framework for azimuthal anisotropy, extended to identified mesons and baryons. Analysis of $v_2(p_T,\text{cent})$ and $v_3(p_T,\text{cent})$ disentangles contributions from initial geometry, radial flow, partonic energy loss, hadronic re-scattering, and baryon number. The scaling functions simultaneously constrain $\eta/s$, $\hat{q}$, radial flow, re-scattering, and EOS stiffness, with dependencies on $\langle N_{\text{chg}} \rangle$, system size, and beam energy. LHC systems (Pb+Pb, Xe+Xe) show low $\eta/s$, strong radial flow, and negligible re-scattering, while Au+Au at RHIC energy exhibits reduced $\eta/s$ and flow with modest re-scattering. Continuity across collective and quenching regimes underscores the universality of the framework and establishes it as a quantitative diagnostic of QGP transport and EOS constraints.

%
\bibliography{pid-refs}
%
\end{document}